\documentclass[journal]{IEEEtran}
\ifCLASSINFOpdf
\else
   \usepackage[dvips]{graphicx}
\fi
\usepackage{url}
\usepackage{graphicx}
\usepackage{cite}
\usepackage{flushend}
\usepackage{color}
\usepackage{mathrsfs}

\begin{document}

\title{Deep Transfer Clustering of Radio Signals}

\author{
    \vskip 1em
    Qi~Xuan,~\emph{Member,~IEEE},
    Xiaohui~Li,
    Zhuangzhi~Chen,
    Dongwei~Xu,
    Shilian~Zheng,
    and~Xiaoniu~Yang
\thanks{This work was supported in part by the National Natural Science Foundation of China under Grants 61973273 and 61903334, and by the Zhejiang Provincial Natural Science Foundation of China under Grants LR19F030001 and LY21F030016. \emph{(Corresponding authors: Qi Xuan.)}}
\thanks{Q. Xuan, X. Li, Z. Chen, and D. Xu are with the Institute of Cyberspace Security, College of Information Engineering, Zhejiang University of Technology, Hangzhou 310023, China (e-mail: xuanqi@zjut.edu.cn).}
\thanks{S. Zheng is with the Science and Technology on Communication Information Security Control Laboratory, Jiaxing 314033, China.}
\thanks{X. Yang is with the Institute of Cyberspace Security, Zhejiang University of Technology, Hangzhou 310023, China, and also with the Science and Technology on Communication Information Security Control Laboratory, Jiaxing 314033, China.}
}
\maketitle

\begin{abstract}
Modulation recognition is an important task in radio signal processing. Most of the current researches focus on supervised learning. However, in many real scenarios, it is difficult and cost to obtain the labels of signals. In this letter, we turn to the more challenging problem: can we cluster the modulation types just based on a large number of unlabeled radio signals? If this problem can be solved, we then can also recognize modulation types by manually labeling a very small number of samples. To answer this problem, we propose a deep transfer clustering (DTC) model. DTC naturally integrates feature learning and deep clustering, and further adopts a transfer learning mechanism to improve the feature extraction ability of an embedded convolutional neural network (CNN) model. The experiments validate that our DTC significantly outperforms a number of baselines, achieving the state-of-the-art performance in clustering radio signals for modulation recognition.  
\end{abstract}


\IEEEpeerreviewmaketitle

\section{Introduction}


\IEEEPARstart{W}{ith} the development of radio communication technology, the electromagnetic environment has become increasingly complex, and the amount of radio signals has also exploded. As the basis of radio communication, signal modulation recognition is of particular importance. Recently, many deep learning models have been applied to signal modulation classification. Most of these works focus on supervised learning, which relies on a large number of labeled signals. However, labeling a large number of signals could be difficult and costly in reality. In order to make better use of available unlabeled signals, clustering is a promising direction. As a kind of unsupervised learning method, clustering can directly captures the correlation between signals, so as to group them into multiple clusters without the need for signal labels in advance. However, it is real a challenge to cluster radio signals for modulation recognition since the signal waves of the same modulation type could be quite different, while those of different modulation types may be close to each other, since the difference of signal waves could be largely determined by the transmitted information. To the best of our knowledge, there are few studies on the modulation clustering of radio signals, which is used to analyze the importance of various features in modulation recognition  ~\cite{DALDAL201945}, and to reconstruct the cluster center vector of the constellation diagram~\cite{JAJOO201713, 8594716, 8520746, zhao2020density}.


The researches on clustering are much more active in other areas, such as computer vision~\cite{Tian_Gao_Cui_Chen_Liu_2014,10.1145/3219819.3220068,10.1145/3219819.3220000,10.1007/978-3-030-04167-0_33,572108} and time-series analysis~\cite{1908.05968, 8704258}. Quite recently, a number of deep learning models are proposed for image clustering, which can be roughly divided into two groups: end-to-end methods and two-step methods. For the first group, samples are soft-labeled according to the clustering results to guide the training of deep learning models. For example, deep embedded clustering (DEC)~\cite{pmlr-v48-xieb16} soft-labeled samples based on the Student $t$ distribution, joint unsupervised learning (JULE)~\cite{Yang_2016_CVPR} used $K$-nearest neighbor (KNN), and deep adaptive clustering (DAC)~\cite{Chang_2017_ICCV} is based on the similarity between samples. For the second group, feature learning is separated from the clustering process. Most of these methods use deep learning models such as autoencoders to learn features, and then cluster them, e.g., deep density-based image clustering (DDC)~\cite{REN2020105841}. In this case, the feature learning process is not guided by clustering. Since the two processes are separated, the features learned by the model may not meet the clustering requirements, which may hurt the performance of the methods.


The above deep clustering methods are largely determined by the training of deep learning models, and cannot be directly adopted to realize signal clustering for modulation recognition due to its essential challenge. 
Transfer learning~\cite{yu2007boosting, raina2007self, evgeniou2004regularized, yosinski2014transferable}, on the other hand, is proposed to solve the problem of insufficient samples. It can largely utilize knowledge or patterns learned from a different but related fields or problems. Therefore, it is naturally to believe that the performance of deep clustering methods could be significantly enhanced if we use transfer learning to pre-train the deep learning models based on an auxiliary dataset in the related fields, and then use it to cluster the samples in the target dataset.


\begin{figure*} [!t]
\centering
\includegraphics[width=0.9\textwidth]{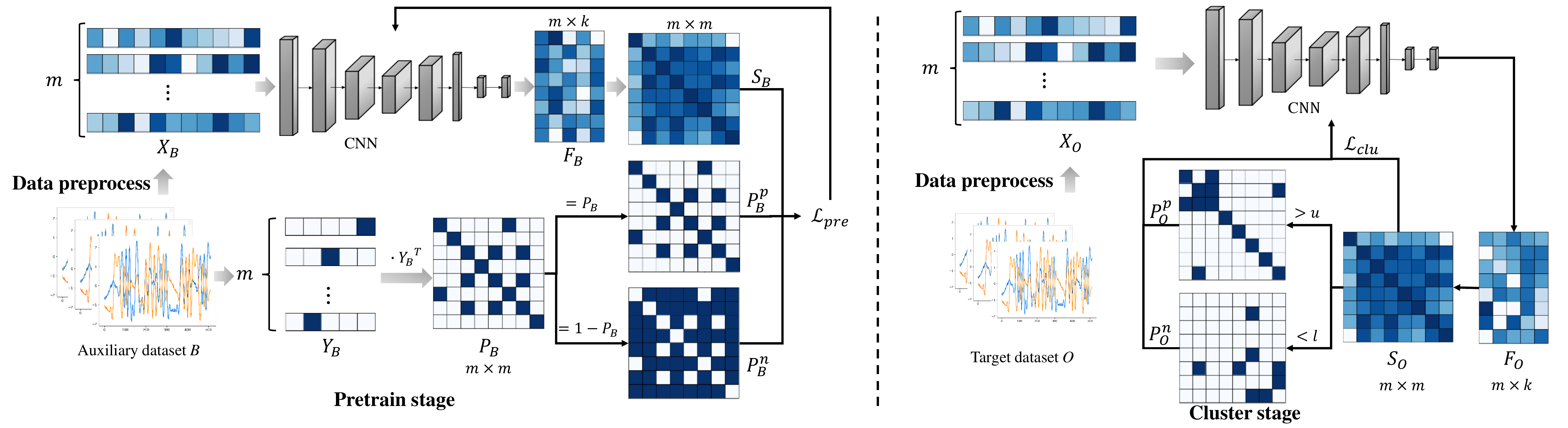}
\caption{The overall framework of DTC for signal clustering, including three stages: data preprocessing, pre-training, fine-tuning and clustering.}
\label{Fig:DTC}
\end{figure*}

In this letter, we propose a deep transfer learning (DTC) of radio signals for modulation recognition for the first time, which naturally integrates feature learning and clustering, and adopts a transfer learning strategy to enhance the feature extraction ability. In particular, firstly, a convolutional neural network (CNN) model is pre-trained with labeled signals of an auxiliary dataset in the same field as the target dataset. In this process, due to the true labels of the signals, the effective feature learning could be guaranteed. After that, iterative cluster training is performed on the target dataset to fine-tune the CNN model. Note that the pre-trained CNN model is deployed to perform preliminary feature extraction of the signals for clustering, while those signals with high confidence of clustering results will be selected and labeled as soft labels to further fine-tune the CNN model. The process is carried out iteratively until the clustering accuracy no longer improves. Since each time the signals with high confidence are used for the model training, the model has a certain degree of anti-interference ability, and at the same time, the training efficiency could be improved. Since the clustering process and the feature learning process are jointly trained, this method can better obtain hidden features that are more suitable for signal clustering. The main contributions of this letter are summarized as follows:

\begin{enumerate}
\item We propose a deep transfer clustering (DTC) model for radio signals, which naturally integrates feature learning and deep clustering for the first time in this area.
\item We adopt a transfer learning mechanism for the supervised pre-training of the CNN model, which effectively improves the feature extraction ability of our DTC model for signal clustering. 
\item We create a loss function consisting of two parts: positive loss and negative loss, the balance between which is adjusted by a hyperparameter $\lambda$.
\item Experimental results validate that our DTC model significantly outperform the other traditional or deep learning based clustering methods on multiple radio signal datasets, achieving the state-of-the-art performance.
\end{enumerate}
The rest of paper is organized as follows. In Section~\ref{sec:method}, we introduce our DTC model in detail, including data preprocessing, pre-training, fine-tuning and clustering. In Section~\ref{sec:exp}, we give the experimental results on three public radio signal datasets, to validate the effectiveness of the DTC model. Finally, the paper is concluded in Section~\ref{sec:conclusion}.

\section{Deep transfer clustering}
\label{sec:method}
In this section, we introduce the detail of our method. The overall framework of DTC is shown in Fig.~\ref{Fig:DTC}, which includes three stages: data preprocessing, pre-training, and clustering. 

\subsection{Data preprocessing}
The target dataset $O$ needs to be clustered into $k$ classes. The auxiliary dataset $B$ is a labeled signal dataset that is independent from $O$. We will use $B$ to pre-train the model before clustering $O$. Since $O$ and $B$ are independent from each other, the signals from these two sets may be of different length. Then, the length of signals in $B$ is first adjusted to be the same as $O$. When the signals in $B$ are shorter than those in $O$, they are expanded to the target length by just copying. For example, signal "$abb$" of length three is expanded to "$abbabbab$" of length eight. When signals in $B$ are longer than those in $O$, they are compressed based on equal-interval sampling to keep the structural characteristics of signals. 
Of course, the number of categories in the two datasets may not be the same. When the number of sample categories in $B$ is more than $k$, we only select the $k$ category samples to use, here we choose randomly; When the number of categories is not enough, all samples are used, but the effect of pre-training process will be slightly reduced.

\subsection{Pre-training}
In order to improve the feature extraction ability of the convolutional neural network (CNN), the signals in $B$, including their labels, are then used to pre-train the model. 

We randomly select $m$ signals from dataset $B$ as the batch input of the CNN, and the output is an $m\times{k}$ signal feature matrix $F_B=\{f_i\}^m_{i=1}$, where $f_i$ is the feature vector of signal $i$ in $B$, with feature dimension equal to $k$. The cosine similarity between signals $i$ and $j$ is defined as:
\begin{eqnarray}
sim(x_i, x_j)=\frac{f_i \cdot f_j}{\left\|f_i\right\| \cdot \left\|f_j\right\| },
\end{eqnarray}
which is simplified to 
\begin{eqnarray}
sim(x_i, x_j)=f_i \cdot f_j,
\end{eqnarray}
when we set $\left\|f_i\right\|=1$ for $i=1,2,\cdots,m$.
So the similarity matrix of these $m$ signals is
\begin{eqnarray}
S_B=F_B \cdot F_B^\mathrm{T}.
\end{eqnarray}

The labels of these signals are converted into one-hot vectors of length $k$, which are grouped into an $m\times{k}$ label matrix $Y_B=\{y_i\}^m_{i=1}$. Then the true binary judgment matrix of the signals is defined as:
\begin{eqnarray}
P_B=Y_B \cdot Y_B^\mathrm{T},
\end{eqnarray}
which is a Boolean matrix, with its element $P_B(i,j)=1$ if signals $i$ and $j$ belong to the same category, and $P_B(i,j)=0$ otherwise. Based on $P_B$, we define a positive matrix $P_B^p=P_B$ and a negative matrix $P_B^n=1-P_B$. Then, the loss function in the pre-training process is defined as:
\begin{eqnarray}
\mathcal{L}_{pre}=-P_B^p \cdot \log{S_B} - \lambda P_B^n \cdot \log{(1-S_B)}, \label{Loss1}
\end{eqnarray}
where $\lambda$, as a hyperparameter, is used to adjust the proportion of positive losses and negative losses. 

The pre-training process stops when the loss value on the validation set of $B$ no longer drops, and we think the CNN has a good ability of feature extraction. 

\subsection{Fine-tuning and clustering}
Now, the target dataset $O$ is also divided into batches as the input of the pre-trained CNN, with the batch size is set to $m$, as shown in the right of Fig.~\ref{Fig:DTC}. For each batch input, we have the output feature matrix $F_O$, also we can obtain the similarity matrix 
\begin{eqnarray}
S_O=F_O \cdot F_O^\mathrm{T}.
\end{eqnarray}
The elements of $S_O$ are then compared with the upper threshold $u$ and lower threshold $l$, respectively, to determine whether the corresponding signals belong to the same cluster or not. 

We construct the positive matrix $P_O^p$ and the negative matrix $P_O^n$ with their elements defined as
\begin{eqnarray}
P_O^p(i,j)=
\left\{
             \begin{array}{lr}
             1, \quad if \ S_O(i,j) \geq u  &  \\
             0, \quad if \ S_O(i,j) < u &\\
             \end{array} i,j = 1, \dots , m 
\right.
\end{eqnarray}
\begin{eqnarray}
P_O^n(i,j)=
\left\{
             \begin{array}{lr}
             1, \quad if \ S_O(i,j) \leq l  &  \\
             0, \quad if \ S_O(i,j) > l &\\
             \end{array} i,j = 1, \dots , m  
\right.
\end{eqnarray}
Then, it is considered that signals $i$ and $j$ are from the same category if $P_O^p(i,j)=1$, while they belong to different categories if $P_O^n(i,j)=1$. 
In the process of fine-tuning, $P_O^p$ and $P_O^n$ are used as the soft labels to replace the true labels of the signals, and the loss function in the cluster training stage is defined as:
\begin{eqnarray}
\mathcal{L}_{clu}=-P_O^p \cdot \log{S_O} - \lambda P_O^n \cdot \log{(1-S_O)}. \label{Loss2}
\end{eqnarray}


During the training process, the feature vectors of signals from different categories tend to be perpendicular to each other. Note that the dimension of feature vector is set to $k$, which is the same as the number of categories, the feature vector is normalized, and each element is limited between $0$ and $1$. Therefore, as the training progresses, the output features tend to be in the form of one-hot vectors. The characteristics of the output actually represent the probability distribution of the signals in each category. In other words, the index of the maximum value of the feature vector can be directly used as the label of the signal.

\begin{figure}
\centerline{\includegraphics[width=0.45\textwidth]{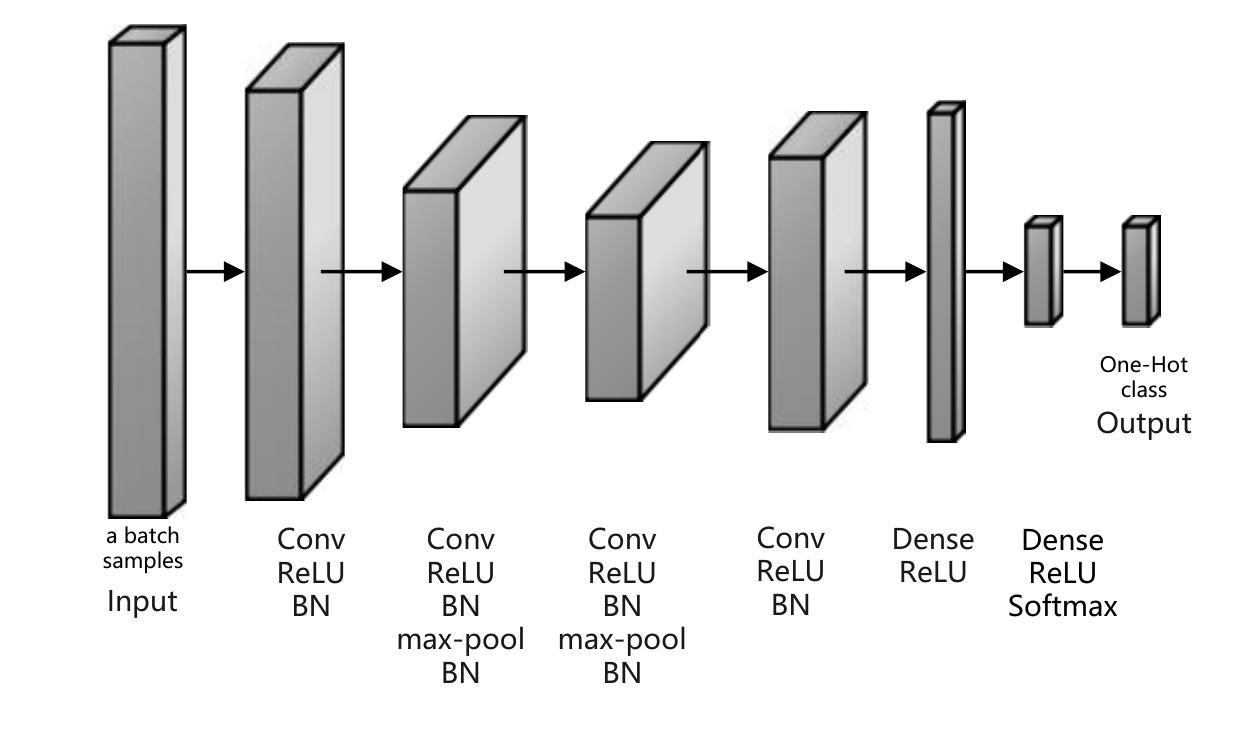}}
\caption{The structure of CNN.}
\label{Fig:CNN}
\end{figure}

In particular, our CNN consists of four convolutional layers and two dense fully connected layers. Each layer use rectified linear (ReLU) activation function. In order to prevent over-fitting, batch normalization (BN) is added before the ReLU layers. At the same time, BN layers can adjust the distribution of the data to a normal distribution to ensure the generalization performance of the model when the input distribution is different at each time. To remove redundant information, the max-pooling layers are added after the ReLU layers of the second and third convolutional layers, and the outputs are also been adjusted by the BN layers. The illustration of the CNN architecture is shown in Fig.~\ref{Fig:CNN}. The model contains 32, 128, 128, and 32 filters in layers 1 to 4, respectively. And the last two dense layers contain 64 and $k$ neurons, respectively. At the end of the model is the softmax function, which acts as a classifier and outputs the probability distribution.

The training of the model uses the Adam optimizer and the loss functions at different stages are defined by Eq.~(\ref{Loss1}) and Eq.~(\ref{Loss2}), respectively. All experiments are run on NDIDIA Tesla V100 based on the TensorFlow deep learning framework.

\begin{table*}[htbp] 
\centering 
\caption{The modulation types of the three datasets}
\label{tab:types}
\begin{tabular}{ccc}
\hline 
\hline 
Datasets & Modulation Types\\
\hline 
RML2016.10A & WBFM,QPSK,64QAM,16QAM,4PAM,GFSK,CPFSK,BPSK,8PSK,AM-SSB \\
RML2016.04C & WBFM,QPSK,64QAM,16QAM,4PAM,GFSK,CPFSK,BPSK,8PSK,AM-SSB \\
RML2018.01A & 32PSK,16APSK,32QAM,FM,GMSK,32APSK,OQPSK,8ASK,16PSK,64APSK \\
\hline 
\hline 
\end{tabular}
\end{table*}

\begin{table*}[htbp] 
\centering 
\caption{The clustering results of cross pre-training on the three datasets}
\label{table1}
\begin{tabular}{ccccccccccc}
\hline 
\hline 
& Datasets & \multicolumn{3}{c}{RML2016.10A} & \multicolumn{3}{c}{RML2016.04C} & \multicolumn{3}{c}{RML2018.01A} \\
\hline 
& Metrics & NMI & ARI & ACC & NMI & ARI & ACC & NMI & ARI & ACC \\ 
\hline 
& No pre-train & 0.3309 & 0.2372 & 0.3699 & 0.4336 & 0.2662 & 0.3905 & 0.3898 & 0.2205 & 0.3082 \\
Auxiliary & RML2016.10A & —— & —— & —— & \textbf{0.8259} & \textbf{0.6888} & \textbf{0.7137} & \textbf{0.6576} & \textbf{0.4516} & \textbf{0.4831} \\
dataset& RML2016.04C & \textbf{0.8547} & \textbf{0.7566} & \textbf{0.7444} & —— & —— & —— & 0.6674 & 0.4587 & 0.4321 \\
& RML2018.01A & 0.5441 & 0.3716 & 0.4980 & 0.6768 & 0.5213 & 0.6229 & —— & —— & —— \\
\hline 
\hline 
\end{tabular}
\end{table*}

\begin{table*}[htbp] 
\centering 
\caption{The clustering results of various methods on the three datasets}
\label{table2}
\begin{tabular}{cccccccccc}
\hline 
\hline 
Datasets & \multicolumn{3}{c}{RML2016.10A} & \multicolumn{3}{c}{RML2016.04C} & \multicolumn{3}{c}{RML2018.01A} \\
\hline 
Metrics & NMI & ARI & ACC & NMI & ARI & ACC & NMI & ARI & ACC \\ 
\hline 
K-means & 0.1345 & 0.0585 & 0.1946 & 0.2674 & 0.1494 & 0.3186 & 0.3573 & 0.0933 & 0.1355 \\
DEC & 0.1150 & 0.0626 & 0.2160 & 0.3034 & 0.2022 & 0.3865 & 0.2741 & 0.0887 & 0.1758 \\
DAC & 0.3081 & 0.2345 & 0.3616 &  0.4707 & 0.2880 & 0.3984 & 0.3768 & 0.2174 & 0.3067 \\
DTC & \textbf{0.8547} & \textbf{0.7566} & \textbf{0.7444} & \textbf{0.8259} & \textbf{0.6888} & \textbf{0.7137} & \textbf{0.6576} & \textbf{0.4516} & \textbf{0.4831}  \\
\hline 
\hline 
\end{tabular} 
\end{table*}


\section{Experiments}\label{sec:exp}
\subsection{Datasets}
The experiments are conducted on three publicly available datasets~\cite{chen2020signet}, including RML2016.10A, RML2016.04C, and RML2018.01A. However, in our experiments, only 10 categories of signals in each dataset are used. We set the modulation types in RML2016.10A and RML2016.04C the same, while they are totally different from those in RML2018.01A, as presented in Table~\ref{tab:types}.  

\subsection{Experimental settings}
\begin{itemize}
    \item \textbf{Baselines}: we compare the proposed DTC with several existing clustering methods, including K-means, DEC, and DAC. The codes of DEC and DAC used in the experiments are downloaded from GitHub, with the parameters set as suggested and the signals reshaped as required.
    \item \textbf{Evaluation metrics}: we use three popular metrics, including adjusted rand index (ARI), normalized mutual information (NMI), and clustering accuracy (ACC), with their values all in [0, 1] and higher scores indicating better clustering performance.
    \item \textbf{Hyperparameters}: we set $\lambda=0.1$ for pre-training and $\lambda=100$ for fine-tuning and clustering, and set the upper threshold $u=0.95$, and the lower threshold $l=0.7$.
\end{itemize}




\subsection{The experimental results of DTC}
The experimental results of DTC are shown in Table~\ref{table1}, where we can see that DTC can achieve reasonable results even without pre-training, i.e., both NMI and ACC are above 0.3, while ARI is above 0.2 for all the three datasets. Meanwhile, the performance of DTC is indeed significantly boosted when the CNN model is pre-trained by auxiliary dataset. All the clustering results on any evaluation metric are significantly improved, no matter which auxiliary dataset is used to pre-train the model for which target dataset. Taking the RML2016.10A dataset as an example, without pre-training, NMI, ARI, and ACC are 0.3309, 0.2372, and 0.3699, respectively. However, after pre-training on the dataset RML2016.04C, these three metrics greatly increase to 0.8547, 0.7566, and 0.7444, respectively. Note that the two datasets RML2016.10A and RML2016.04C are quite similar to each other, i.e., they share exactly the same types of modulation and the same length of signals, while the dataset RML2018.01A is relatively different on both types of modulation and length of signals. As expected, the DTC for RML2016.10A benefits most from the CNN model pretrained on RML2016.04C, and vice versa. More interestingly, though quite different, the CNN models pretrained by RML2018.01A can still help to extract important features of the signals in the other two datasets, so as to improve the performance of DTC, which indicates the generalization ability of our method.

\subsection{Comparison with other clustering methods}
Now, we compare our DTC with other clustering methods, including K-means, DEC, and DAC. K-means is a very popular clustering method in many areas, DEC and DAC are two typical deep clustering methods with outstanding performance in computer vision. Note that we also try several latest deep clustering methods, such as semantic pseudo-labeling for image clustering (SPICE)~\cite{niu2021spice}, robust learning for unsupervised clustering (RUC)~\cite{Park_2021_CVPR}, and semantic clustering by adopting nearest neighbors (SCAN)~\cite{10.1007/978-3-030-58607-2_16}, but the results are worse than the three baselines we choose. The comparison results are shown in Table~\ref{table2}, where we can see that DTC significantly outperforms all the other clustering methods, achieving the state-of-the-art performance. In particular, the clustering accuracy of DTC is as high as 0.7444 on RML2016.10A, which is 105.9\% higher than the second best method DAC. On RML2016.04C, the clustering accuracy of DTC is 0.7137, which is 79.1\% higher than the second best method DAC. On RML2018.01A, these numbers are 0.4831 and 57.5\%. Such incredible results suggest that DTC could be a feasible method to cluster radio signals for modulation recognition as a challenging task in wireless communication. 


\section{Conclusion}
\label{sec:conclusion}
Automatic modulation recognition is crucial for many applications in electromagnetic space, especially when the 5G/6G wireless systems emerge. However, it is always difficult to label a large number of radio signals in many real scenarios, making it hard to use supervised learning to recognize modulation types. Therefore, in this letter, we focus on clustering radio signals for  modulation recognition. Since the signal waves could be largely determined by the transmitted information, the observed signals of the same modulation type could be quite different, while those of different modulation types may be close to each other. This makes clustering radio signals real a challenge in reality.   

With the help of the strong feature extraction ability of convolutional neural network (CNN), in this letter, we propose a novel end-to-end deep transfer clustering (DTC) model for radio signals, which naturally integrates deep learning and transfer learning into a single framework to improve the clustering performance. The experimental results show that, compared with a number of baselines, our method achieves significantly better performance on three public radio signal datasets. In the future, we will apply our DTC model on more various signal datasets, to validate its generalization ability more comprehensively.

\bibliographystyle{ieeetr}
\bibliography{reference.bib}

\end{document}